\documentclass[%
 aip,
 amsmath,
 amssymb,
 superscriptaddress,
 reprint,%
]{revtex4-1}

\usepackage{graphicx}
\usepackage{dcolumn}
\usepackage{bm}
\usepackage{upgreek}

\usepackage{xcolor}

\usepackage{hyperref}

\makeatletter
\newcommand{\manuallabel}[2]{\def\@currentlabel{#2}\label{#1}}
\makeatother

\begin{document}
\manuallabel{fig_saeulen}{S3}
\manuallabel{fig_abs}{S2}
\manuallabel{fig_qwp_charac}{S1}

\preprint{AIP/123-QED}

\title[Photocurrent measurements in topological insulator $\text{Bi}_2\text{Se}_3$ nanowires]{Photocurrent measurements in topological insulator $\text{Bi}_2\text{Se}_3$ nanowires}

\author{N. Meyer}
\email{nina.meyer@uni-greifswald.de}
\thanks{corresponding author}
\affiliation{Institute of Physics, University of Greifswald, Felix-Hausdorff-Str. 6, 17489 Greifswald, Germany}
\author{K. Geishendorf}
\affiliation{IFW Dresden, Institute for Metallic Materials, Helmholtzstra\ss e 20, 01069 Dresden, Germany}
\author{J. Walowski}
\affiliation{Institute of Physics, University of Greifswald, Felix-Hausdorff-Str. 6, 17489 Greifswald, Germany}
\author{A. Thomas}
\affiliation{IFW Dresden, Institute for Metallic Materials, Helmholtzstra\ss e 20, 01069 Dresden, Germany}
\author{M. M{\"u}nzenberg}
\affiliation{Institute of Physics, University of Greifswald, Felix-Hausdorff-Str. 6, 17489 Greifswald, Germany}

\date{\today}

\begin{abstract}
Circular photogalvanic currents are a promising approach for spin-optoelectronics. To date, such currents have  been induced in topological insulator flakes or extended films. It is not clear whether they can be generated in nanodevices. In this paper, we demonstrate the generation of circular photogalvanic currents in $\text{Bi}_2\text{Se}_3$ nanowires. Each nanowire shows topological surface states. Here, we generate and distinguish the different photocurrent contributions via the driving light wave. We separate the circular photogalvanic currents from those due to thermal Seebeck effects, through controlling the laser light polarization. The results reveal a spin-polarized surface-Dirac electron flow in the nanowires arising from spin-momentum locking and spin-orbit effects. The second photocurrent contribution described in this letter is caused by the thermal Seebeck effect. By scanning the photocurrent, it can be spatially resolved; upon reversing the gradient direction along the nanowire, the photocurrent changes its sign, and close to the gold contacts, the amplitudes of the different photocurrent contributions are affected by the proximity to the contacts. In the center of the nanowires, where the effects from the gold contact/ topological insulator stacks vanish, the spin-polarized current remains constant along the nanowires. This allows the all-optical spin current generation in topological insulator nanowires and hybrid structures for nanoscale, one goal of spin-orbitronics. 

\end{abstract}

\maketitle

\begin{figure}
\includegraphics[scale=0.75]{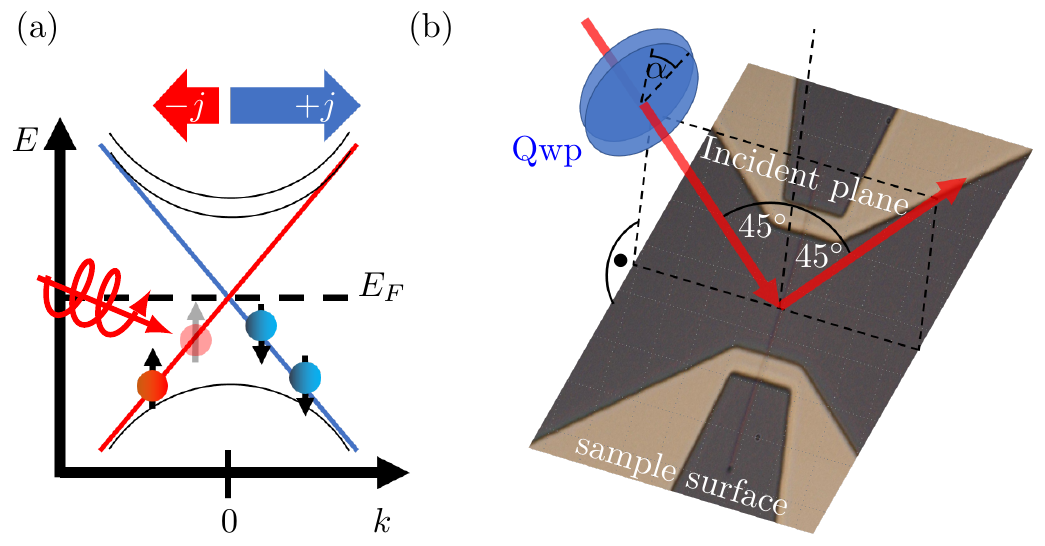}
\caption{\label{fig_skizze}(a) depicts a schematic band structure of $\text{Bi}_2\text{Se}_3$ in the presence of the circular photogalvanic effect. The asymmetric population of the surface states (implied by the unequal number of red and blue dots) is generated by the absorbed circular polarized light (red arrow) and generates a net spin-polarized electrical current due to the spin momentum locking. In (b), the setup geometry is depicted. The red arrows represent the incoming and reflected beam which hit the sample surface under an angle of incident of $45^\circ$. The polarization of the exciting light changes with the angle $\alpha$ of the qwp.}
\end{figure}

In the last decade, a class of topological matter, the so-called quantum spin Hall (QSH) insulators or topological insulators (TIs), theoretically predicted in 2005 by Kane and Mele and realized in 2D in 2007 by K{\"o}nig et al. in CdTe/HgTe quantum wells \cite{KaneMele2005, Koenig2007}, has attracted great attention. This class of matter has the interesting property of an energy gap between the valence and conduction bands that is closed at the boundaries by gapless surface states with high mobility and strong spin polarization \cite{HasanKane2010}. The surface states are predicted to be topologically protected, since their origin lies in intrinsic bulk properties such as large spin-orbit coupling, opposite parity of the bulk bands, band inversion and time-reversal symmetry, which suppresses backscattering and thereby decreases the sensitivity to surface impurities or defects. Additionally, the direction of the spin and momentum for the surface states are locked, giving rise to a strong spin polarization that makes these materials very interesting for spintronic applications e.g. as spin injectors \cite{Huang2017}.\newline After the realization of a 1D topologically non-trivial edge state, 3D topological insulators  consisting of $\text{Bi}_{1-x}\text{Sb}_x$, $\text{Bi}_2\text{Se}_3$, $\text{Bi}_2\text{Te}_3$ and $\text{Sb}_2\text{Te}_3$ were also realized\cite{Plucinski2011, Biswas2015}. Over the last decade, the electronic properties of $\text{Bi}_2\text{Se}_3$ and $\text{Bi}_2\text{Te}_3$ have been extensively studied, with optimization of the properties of their topologically protected surface states. This has been carried out mainly by investigating the band structure of the surface electrons by imaging the band structure using angle resolved photoemission spectroscopy (ARPES) \cite{Krumrain2011, Pan2011},  including the spin-resolved variant, and  time-resolved two-photon photoelectron (2PPE) spectroscopy \cite{Sobota2014}. Scanning tunneling microscopy (STM) can also be used to prove the existence of a linear dispersion relation and suppressed backscattering \cite{Romanowich2013}.\newline To make use of these surface states in spintronics it is essential not only to demonstrate the existence of spin-polarized surface states but also to control them. One approach for controlling these surface states is to create an asymmetrically populated Dirac cone, which leads to spin-polarized currents on the surface due to spin-momentum locking. This idea has been realized among others for exfoliated $\text{Bi}_2\text{Se}_3$ \cite{McIver2011} and other materials and is enabled by the strong spin-orbit coupling in topological insulators such as $\text{Bi}_2\text{Se}_3$ lifting the spin degeneracy of the electrons in the surface states. As a result, the selection rules for interband transitions depend on the electron spin. Therefore, it is possible to use circular polarized light to selective excite surface electrons with a parallel or antiparallel spin component with respect to the photon momentum and depending on the helicity. This creates an asymmetric population of the surface states in k-space, which due to the spin-momentum locking, leads to a spin-polarized electrical current (see FIG.\,\ref{fig_skizze}~(a)). This effect of generating a spin-polarized charge current by creating an asymmetric carrier population in k-space by exciting optical transitions with circular polarized light imposed by optical selection rules is referred to as the circular photogalvanic effect (CPGE) and has been previously realized in semiconductor quantum wells \cite{Ganichev2003}.\newline More recently several studies have performed photocurrent measurements of the $\text{Bi}_{1-x}\text{Sb}_x$, $\text{Bi}_2\text{Se}_3$, $\text{Bi}_2\text{Te}_3$ and $\text{Sb}_2\text{Te}_3$ group using photon wavelengths ranging from the visible to the terahertz to the infrared regime \cite{Kastl2015, Okada2016, Plank2016, Braun2016, Seifert2017}. In this work, we demonstrate spin-polarized photogalvanic effects in $\text{Bi}_2\text{Se}_3$ nanowires with a cross-section in the nanometer range instead of in the micrometer-wide Hall bar structures used in previous experiments \cite{McIver2011, Thomas2018}. From a geometrical point of view the ratio of surface to bulk states in nanowires should be higher due to their more rectangular cross section which might lead to an increased ratio of spin-polarized currents to photocurrents originating from the bulk states.
\medskip

\begin{figure}
\includegraphics[scale=0.9]{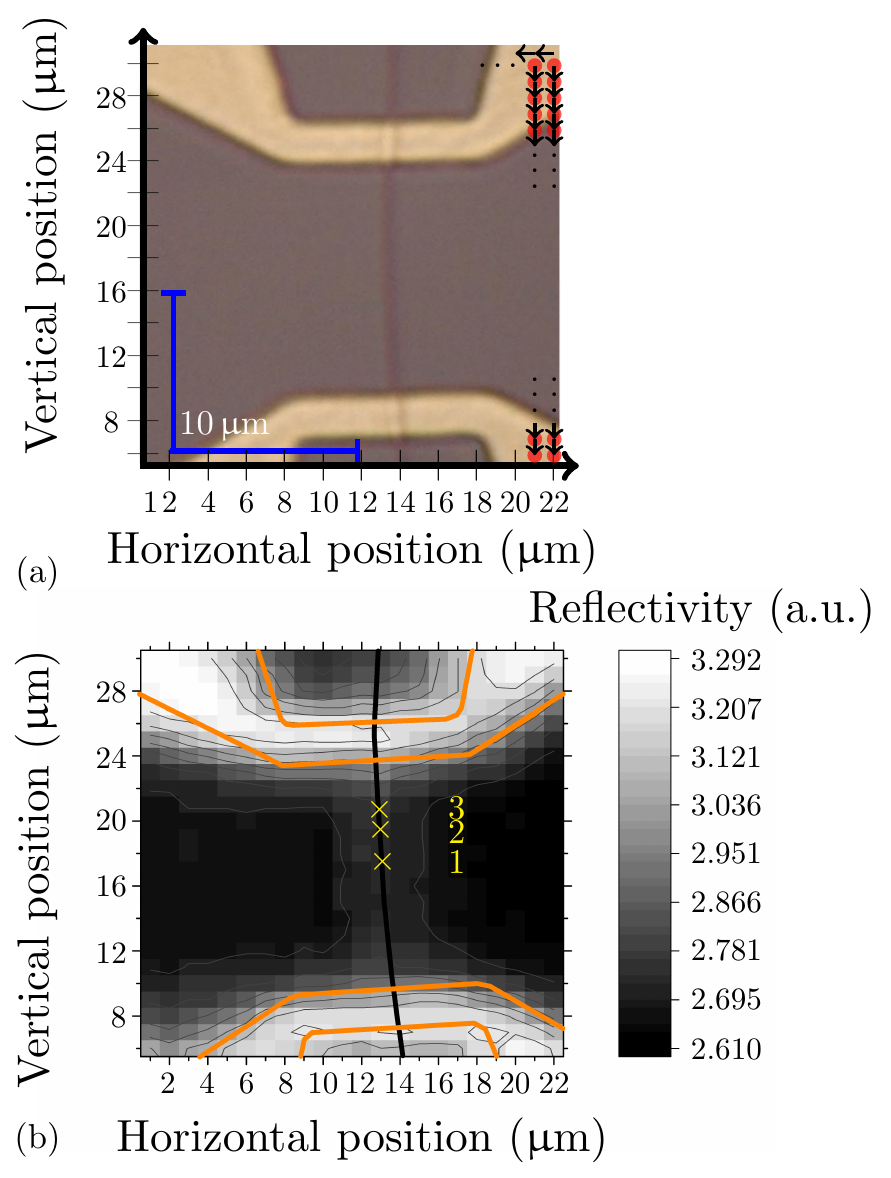}
\caption{\label{fig_measurement1}(a) Shows a light microscopy image of the used sample. The nanowire (dark gray vertical line) is placed on a Si (111) substrate (gray). The gold contacts (yellow) are on top of the nanowire. The two closer parallel edges of the gold contacts are separated by $14\,\mathrm{\upmu m}$. The red dots mark the center of the laser spot on the sample surface. The actual FWHM of the laser spot is four times larger than the dots.The black arrows indicate the direction of movement of the laser spot. The raster pattern starts at (22,30) and ends at (1,6). (b) The edges of the gold contacts (orange) and the position of the nanowire (black line) are marked on the spatially resolved reflected light intensity. Three positions on the nanowire are marked and numbered by 1,2 and 3 (yellow).}
\end{figure}

The $\text{Bi}_2\text{Se}_3$ nanowires were synthesized by Au-catalyzed vapor-liquid-solid method on a Si(111) substrate (as described here: \cite{Shin2016}). The cross-section of the nanowires is either rectangular or trapezoidal, resulting from the layered crystal structure. The width of the nanowires is on the scale of $50\,\mathrm{nm}$ and the thickness is in the range of $50\,\mathrm{nm}-150\,\mathrm{nm}$. This length scales are above $10\,\mathrm{nm}$, where hybridization of the topological surface states followed by a gap opening at the Dirac point appears \cite{Gooth2014, Shin2016}. The length can be as long as several tens of micrometers. The nanowire presented in FIG. \ref{fig_measurement1}~(a) has a total length of $36\,\mathrm{\upmu m}$. The $\text{Bi}_2\text{Se}_3$ nanowires are grown in the [110] direction as a single-crystal structure and have a smooth surface. The chemical composition can be characterized by an energy dispersive spectrometer in the scanning TEM mode. The measured ratio is 2:3 as expected for $\text{Bi}_2\text{Se}_3$. The grown nanowires are mechanically transferred onto a Si (111) substrate, and the gold contacts separated by a $14\,\mathrm{\upmu m}$ gap are fabricated on top of the nanowire by lithography. FIG. \ref{fig_measurement1}~(a) shows a micrograph of one of the nanowire devices with two contacts, which are each connected to two gold pads on the left- and on the right-hand sides of the nanowire \cite{Hamdou2013}. The sample is mounted with silver paste onto a chip carrier and connected with $25\,\mathrm{nm}$ diameter gold wires  by wire bonding. \newline The light source for the photocurrent measurements is a diode laser with a wavelength of $785\,\mathrm{nm}$ $\left(1.55\,\text{eV}\right)$ modulated at a frequency of $77\,\mathrm{Hz}$. The laser light passes through a linear polarizer and a quarter-wave plate (qwp) prior to impinging on the sample surface under an angle of incidence of $45^\circ$. The rotation angle $\alpha$ of the qwp is controlled by a step motor to change the polarization of the excitation beam see FIG.~\ref{fig_skizze}(b). The laser light is focused down to $(4.3\pm0.11)\,\mathrm{\upmu m}\times (2.89\pm0.08)\,\mathrm{\upmu m}$ on the sample surface. The intensity of the light reflected from the sample surface is measured by a photodiode. The photocurrent between the two contacts and the light reflected from the sample surface are simultaneously measured by a lock-in amplifier. The laser spot can be moved across the sample surface along the vertical and horizontal directions by two step motors with a minimum step size smaller than $1\mathrm{\upmu m}$ with an error of $200\,\mathrm{nm}$. The raster pattern of the laser spot (red dots) is depicted in FIG.~\ref{fig_measurement1}~(a), as drawn on a light microscopy image of the sample. The measurement starts with the laser spot in the right upper corner (position (22,30) in FIG.~\ref{fig_measurement1}(a)). Then the qwp is rotated by $\Delta\alpha=6^\circ$ steps to carry out a full rotation while the photovoltage and the intensity of the reflected beam are measured. Afterwards, the laser spot is moved by down by $1\,\mathrm{\upmu m}$ to the position (22,29) as depicted by the raster pattern in FIG.~\ref{fig_measurement1}(a), repeating the measurement procedure until the bottom left position is reached.

\begin{figure}
\includegraphics[scale=0.9]{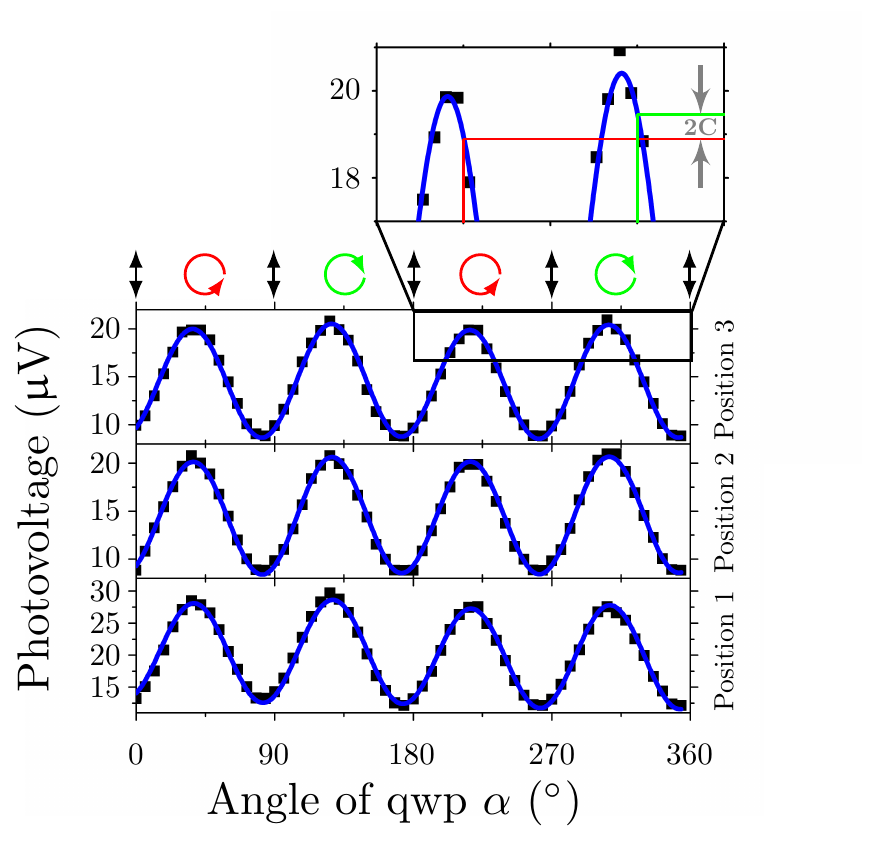}
\caption{\label{fig_bsp_photovoltage} The measured photovoltage (black squares) and the analysis results using eq.~\eqref{eq_current} (blue line) as a function of the rotation angle $\alpha$ for three laser spot positions (see FIG.\ref{fig_measurement1}~(b)) are displayed. In addition, the photovoltages for left and right circular polarized light (indicated by green and red lines) are marked. The difference between these photovoltage values is equal to 2C. The results for the parameter in Eq.~\eqref{eq_current} are shown in FIG.~\ref{fig_saeulen} in the supplementary material.}
\end{figure}
\begin{figure}
\includegraphics[scale=0.9]{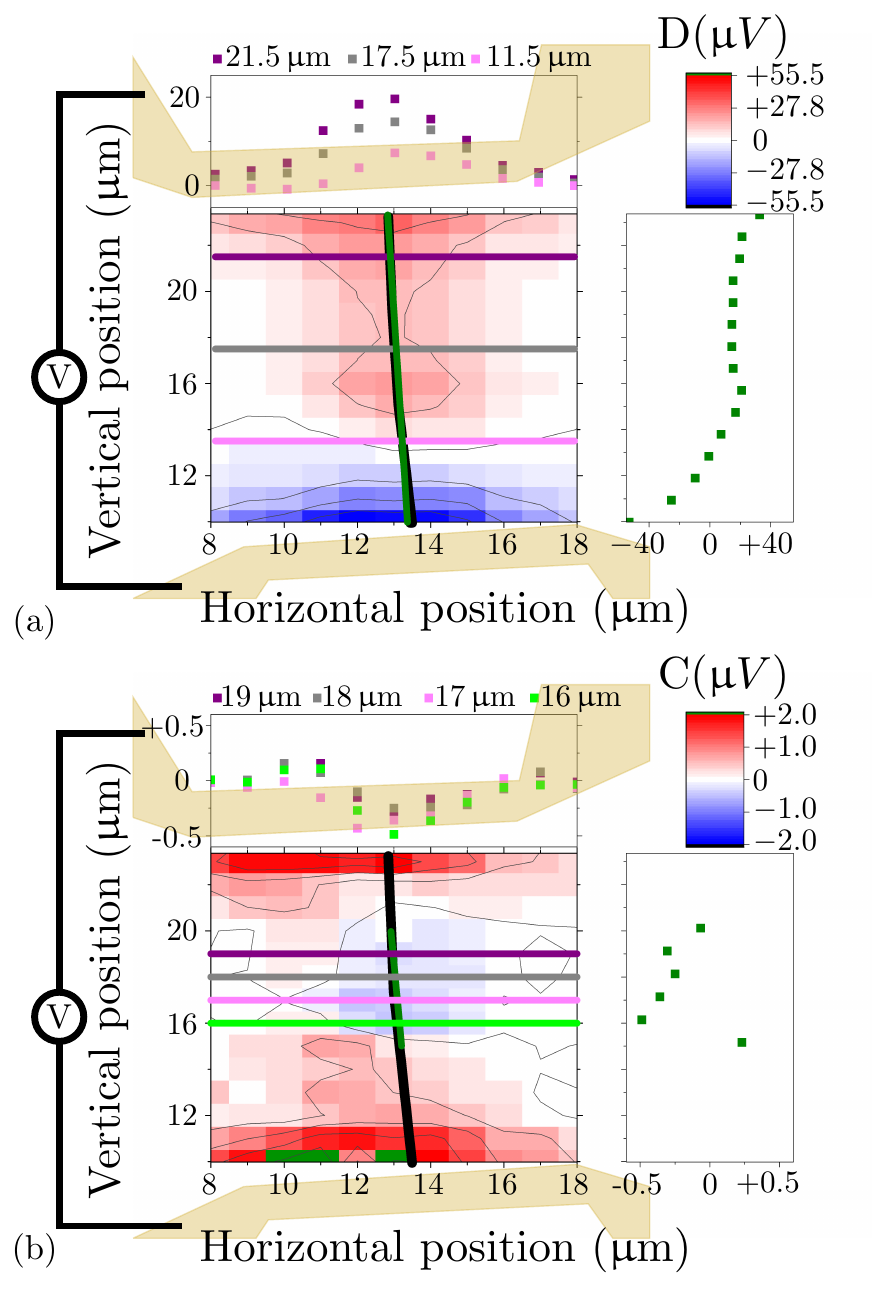}
\caption{\label{fig_measurement2}(a) The map shows the amplitude D related to the thermoelectric voltage at different laser spot positions in the $14\,\mathrm{\upmu m}$ gap between the gold contacts. The vertical contour plot shows the amplitude D along the nanowire position and the horizontal contour plots show how the thermoelectric voltage drops when the center of the laser spot is moved away from the nanowire. (b) The map shows the spatially resolved amplitude C, related to the spin-polarized current. The vertical contour plot shows the spin-polarized voltage along the nanowire (green line) and the four horizontal contour plots prove that the spin-polarized voltage increases when the nanowire is illuminated.}
\end{figure}

For each data point in the two-dimensional voltage maps, the photovoltage between the two contacts is measured while the qwp is rotated by $360^\circ$ in total, changing the polarization from linear polarized ($\alpha=0^\circ$), to left-circular polarized ($\alpha=45^\circ$), to linear polarized ($\alpha=90^\circ$), to right-circular polarized ($\alpha=135^\circ$), to linear polarized ($\alpha=180^\circ$), to left-circular polarized ($\alpha=225^\circ$) and so on and so forth (the characterization of the qwp is in the supplementary material in FIG.~\ref{fig_qwp_charac}). The measured photovoltage $v$ is a sum of four contributions 
\begin{equation}
v\left(\alpha\right)=C\sin(2\alpha)+L_1 \sin (4\alpha)+L_2\cos(4\alpha)+D\label{eq_current}.
\end{equation}
The contributions in Eq.~\eqref{eq_current} can be distinguished by their dependence on the polarization of the exciting laser light, like it has been done previously by McIver et al. for photocurrent measurements in exfoliated $\text{Bi}_2\text{Se}_3$ Hall bar devices \cite{McIver2011}. The first term $C\sin(2\alpha)$ of Eq.~\eqref{eq_current} modulates the difference in the photovoltage for left and right circular polarized light and is zero if the exciting light is linear polarized. According to McIver et al. this term describes the amount of spin-polarized voltage generated by the CPGE while previous measurements by Shalygin et al. \cite{Shalygin2007} show, that also the circular photon drag effect can cause a similar voltage in (110)-grown quantum wells. In this paper we relate the first term in Eq.~\eqref{eq_current} to the CPGE following the notation of the recent works \cite{McIver2011, Luo2017}. Therefore, the magnitude of the amplitude C, which is half of the difference between the photovoltage for different helicities (see Fig.~\ref{fig_bsp_photovoltage}), is used as a measure for the size of the spin-polarized voltage. The second term $L_1 \sin (4\alpha)$ and third term $L_2\cos(4\alpha)$ describe the contributions that arise from the linear photogalvanic effect and the photon drag effect \cite{McIver2011, Olbrich2014, Plank2016B}.
As we show below, the last term D is independent of the polarization and arises from the Seebeck effect. Possible contributions of polarization independent photogalvanic and photon drag effects are negligible. The polarization dependent absorption is modulated by the third term (see FIG.~\ref{fig_abs} in the SI). The Seebeck effect is caused by the heating due to the laser spot, which is smaller than the distance between the gold contacts and therefore, creates an overall temperature gradient that can change in direction and size as the laser spot is moved across the sample surface. In the following Eq.~\eqref{eq_current} is the fitfunction and the four amplitudes $C$, $L_1$ , $L_2$ and $D$ are determined by fitting Eq.~\ref{eq_current} to the measured photovoltage at every laser spot position. The resulting fitfunctions and the measured photovoltage for three positions are shown in FIG.~\ref{fig_bsp_photovoltage} and the results for all four parameters are shown in FIG.~\ref{fig_saeulen} in the SI. The horizontal and vertical positions of the laser spot are then used as the spatial coordinates for the extracted amplitudes as in FIG. \ref{fig_measurement2} for the thermoelectric contribution represented by D and the spin-polarized contribution represented by C. At the same time, the intensity of the reflected light for every value of the photovoltage is measured with a second lock-in amplifier. Instead of fitting Eq.~\eqref{eq_current} to the obtained values, we take the value for a fixed polarization at $\alpha=0^\circ$ and again use the position of the laser spot as the coordinates for the reflectivity to obtain a two-dimensional map of the reflectivity (see FIG.~\ref{fig_measurement1} (b)) which allows us to identify the positions of the nanowires and the gold contacts.

In the reflectivity map (FIG. \ref{fig_measurement1} (b)) three different areas can be distinguished. The area with the highest intensity (white) shows the position of the Au contacts, while the region with the lowest reflectivity (black) represents the $300\,\mathrm{nm}$ SiO on top of the Si substrate.
The nanowire can be clearly distinguished between the contacts (in dark gray). Its size looks exaggerated since its width ($150\,\mathrm{nm}$) is smaller than the spot size of the laser beam, so it acts as a scattering center. Therefore, the reflectivity map proves that the area shown in FIG. \ref{fig_measurement1}~(a) is illuminated and indicates the position of the nanowire. The nanowire is marked in the photovoltage maps (FIG.~\ref{fig_measurement2}) by a black line.\newline The three selected photovoltage measurements in FIG.~\ref{fig_bsp_photovoltage} at three different positions along the nanowire (marked in yellow in FIG. \ref{fig_measurement1}~(b)) show good agreement between Eq.~\eqref{eq_current} and the measured voltage. It is also observed, that the thermoelectric contribution D, which is equal to the shift along the vertical direction, is at least one order of magnitude larger than the spin-polarized contribution C. The spatially resolved maps in FIG.~\ref{fig_measurement2} enable a more detailed inspection of the two contributions. The thermoelectric amplitude D (FIG.~\ref{fig_measurement2}(a)) changes its sign from positive at the top electrode to negative at the lower electrode. Note that the thermoelectric current that creates the thermoelectric voltage is generated by two temperature gradients with opposite signs that point from the laser spot position towards the two colder contacts. Therefore, the position of the laser spot with respect to the contacts determines the size and sign of the thermoelectric voltage. When the laser spot is in the center between the contacts, the two temperature gradients cancel, and thus, the net temperature gradient and D vanish. In our measurements (FIG.~\ref{fig_measurement2}(a)), the direction of the net temperature gradient is encoded in red and blue, corresponding to the gradients pointing towards the lower or upper contacts, respectively. At the vertical position of $13\,\mathrm{\upmu m}$ in FIG. \ref{fig_measurement2}(a), the thermoelectric contribution D becomes zero. Once the laser spot is moved toward one of the electrodes, the net temperature gradient is non-zero, and a net current is generated by the Seebeck effect. The sign of the thermoelectric current changes from top to bottom since the direction of the net temperature gradient is reversed. Thus, the voltage of the contour plot in the vertical direction along the nanowire changes its sign from top to bottom (shown in FIG.~\ref{fig_measurement2} (a)). The contour plots perpendicular to the nanowire reveal that the highest values for D are reached when the center of the laser spot matches the horizontal position of the nanowire at $13\,\mathrm{\upmu m}$. If the laser spot center is moved away from the nanowire along the horizontal direction, then the values decrease. When the laser spot center is $4\,\mathrm{\upmu m}$, which is equal to the full width at half maximum (FWHM) of the laser spot diameter, or further from the nanowire, the thermovoltage drops down to nearly zero. This means that there is no contribution to the thermoelectric contribution D transferred through the substrate. \newline A slight enhancement of the voltage is observed when not only the nanowire but also the nanowire underneath the gold contact is partially illuminated. This appears for the vertical positions above $21\,\mathrm{\upmu m}$ and below $11\,\mathrm{\upmu m}$ and is manifested by the nonlinear increase of the photovoltage in the vertical contour plot and also by the circular shape of the lines in the 2D map in FIG.\ref{fig_measurement2} (a). In addition, the thermoelectric voltage reaches zero at a vertical position of $13\,\mathrm{\upmu m}$, which is $4\,\mathrm{\upmu m}$ away from the geometric center of the nanowire. One possible reason for this shift of the zero-crossing might be the asymmetrical heating of the laser beam, since its gausian heat profile is no longer symmetrical due to the angle of incidence of $45^\circ$. Another reason might be the enhancement of the thermoelectric voltage at the Au/TI layer combined with the laser spot size. The FWHM of the laser spot in vertical direction is $2.89\,\mathrm{\upmu m}$. When the center of the laser is $3\,\mathrm{\upmu m}$ away from the contacts, the Au/TI layer is still heated and contributes to the measured thermoelectric voltage. As a result, the slope towards the lower contact (negative photovoltage) is steeper compared to the slope towards the higher contact (positive photovoltage). In addition, the nanowire in this experiment is slightly bend as shown in FIG.~\ref{fig_measurement2}(a) and the thermoelectric voltage changes drastically, when the laser spot is moved in the horizontal direction. Hence, the center of the laser light might not always match to the center of the nanowire when moved in the vertical direction, leading to small changes of the thermoelectric voltage. This enhancement of the photovoltage when illuminating a  nanowire underneath a metallic contact is also observed for measurements of GaN, ZnO and Si nanowires. In these materials, the observed increase is a result of the Schottky effect \cite{Leonard2015, Park2015, Yeonghwan2005}. In our case, the gold contact is metallic and the TI can act as a semiconducting layer. The sign of the current caused by the band bending at the metal/semiconductor interface changes between the contacts since the band bending is symmetrical with regard to the center of the nanowire; this is in good agreement with the behavior of the thermoelectric voltage D observed in this work. \newline To exclude the influence of the contacts, we focus on the spin-polarized contribution C in this work in the area between $15$ and $20\,\mathrm{\upmu m}$ along the vertical axes displayed in FIG. \ref{fig_measurement2}(b) in green. The contour plots along the horizontal direction show that the spin-polarized voltage C decreases when the center of the laser spot does not match the position of the nanowire at the horizontal position of $13\,\mathrm{\upmu m}$. This proves that the substrate does not contribute to the spin-polarized voltage. The largest value with a modulus of $0.5\,\mathrm{\upmu V}$ is reached when the laser spot center matches the nanowire, which is a factor of 80 smaller than the largest value for the thermoelectric contribution $D=40\,\mathrm{\upmu V}$. The largest values are reached on the small plateau shown in the vertical contour plot over a range of $4\,\mathrm{\upmu m}$. Closer to the contacts, the spin-polarized voltage decreases. 

In summary, we performed photocurrent measurements on $\text{Bi}_2\text{Se}_3$ nanowires and analyzed the spatially resolved results for the spin-polarized and thermoelectric contributions. For the thermoelectric contribution, we observe a sign change of D along and on the nanowire. This indicates that the parameter D is dominated by the Seebeck effect and potential contributions of polarization independent photogalvanic and photon drag effects are negligible. In addition, we detect an enhancement of the thermoelectric and the spin-polarized contribution when the nanowire underneath the gold contacts is illuminated in comparison to illuminating only the TI. We also show that spin-polarized currents can be generated in nanowires within the range of $5\,\upmu m$ along the nanowire by using circular polarized light.\newline Thus, we have demonstrated the ability to drive photogalvanic currents in nanowires, which shows their promising potential for use in photo-spintronics applications in the future.

\section{Supplementary material}
See supplementary material for the characterization of the qwp, more information about the polarization dependent absorption and the results for the  parameters $C$, $L_1$, $L_2$ and $D$ for position 1,2, and  3 in FIG.~\ref{fig_measurement1}.
\section{Acknowledgments}
We are grateful to the German Science Foundation (DFG) for financial support through the priority program SPP1666: `Topological insulators: materials, fundamental properties, devices` (MU1780/10-2).\medskip

This article may be downloaded for personal use only. Any other use requires prior permission of the author and AIP Publishing. This article appeared in Appl. Phys. Lett. 116, 172402 (2020) and may be found at \url{https://doi.org/10.1063/1.5142837}.

\bibliography{aipsamp}

\end{document}